# Measurements of a low temperature mechanical dissipation peak in a single layer of $Ta_2O_5$ doped with $TiO_2$

I Martin[1]*, H Armandula[2], C Comtet[3], M M Fejer[4], A Gretarsson[5], G Harry[6], J Hough[1], J-M M Mackowski[3], I MacLaren[1], C Michel[3], J-L Montorio[3], N Morgado[3], R Nawrodt[7], S Penn[8], S Reid[1], A Remillieux[3], R Route[4], S Rowan[1], C Schwarz[7], P Seidel[7], W Vodel[7], A. Zimmer[7]

[1]SUPA, University of Glasgow, Scotland.

[2]LIGO Laboratory, California Institute of Technology, USA.

[3]Laboratoire des Matériaux Avancés, LMA, CNRS-IN2P3, France.

[4]Stanford University, USA.

[5]Embry-Riddle Aeronautical University, USA.

[6]LIGO Laboratory, Massachusetts Institute of Technology, USA.

[7]Friedrich-Schiller University, Jena, Germany.

[8]Hobart and William Smith Colleges, USA.

Thermal noise arising from mechanical dissipation in oxide coatings is a major limitation to many precision measurement systems, including optical frequency standards, high resolution optical spectroscopy and interferometric gravity wave detectors. Presented here are measurements of dissipation as a function of temperature between 7 K and 290 K in ion-beam sputtered $Ta_2O_5$ doped with $TiO_2$, showing a loss peak at 20 K. Analysis of the peak provides the first evidence of the source of dissipation in doped $Ta_2O_5$ coatings, leading to possibilities for the reduction of thermal noise effects.

*email: i.martin@physics.gla.ac.uk

## 1. Introduction

Fabry-Perot cavities form critical elements in the development of highly frequency-stabilised lasers for high-resolution optical spectroscopy [1], fundamental quantum measurements and optical frequency standards [2, 3] and quantum information science [4]. In addition, current interferometric gravitational wave detectors rely critically on the use of ultra-stable Fabry-Perot (and other) cavity arrangements [5-9]. In all of these systems, recent research has shown that a serious fundamental limit to the inherent performance of the system is set by the Brownian motion associated with the ion-beam-sputtered cavity mirror coatings of amorphous oxide materials [10-13], with this noise source currently limiting, or expected soon to limit, achievable experimental performance.

The magnitude of this Brownian thermal noise is related to the mechanical dissipation factor of the coating materials. It is thus of major importance to determine the exact level of dissipation, and thus thermal noise, expected from specific coatings, understand the mechanism responsible for this dissipation and find methods of minimising it.

Previous studies have shown that in the commonly-used coatings formed from alternating layers of silica ($SiO_2$) and tantala ($Ta_2O_5$), the dissipation is dominated by the $Ta_2O_5$ component [14-16] and can be reduced by doping the $Ta_2O_5$ with $TiO_2$ [17,18], although the mechanism responsible for the dissipation is not yet well understood.

In general, a direct reduction in thermal noise is expected on lowering the operating temperature of materials [19]. A further reduction in thermal noise is possible in materials in which the mechanical dissipation also decreases with temperature [20]. Silicon is one such material, and has been proposed for use as a mirror substrate for future low temperature gravitational wave detectors [21]. However the nature of the variation of mechanical dissipation with temperature depends on the exact mechanism responsible for the dissipation and in some materials the dissipation increases as temperature is decreased from room temperature [22].

One well-known example of a material exhibiting this property is fused silica [23, 24], in which there is a broad peak in mechanical dissipation centered around approximately 40-60 K. This peak is thought to arise from energy dissipation by thermally activated transitions of the oxygen atoms between two energy states in the amorphous $SiO_2$ network [25]. The broad nature of this peak has been associated with the distribution of bond angles in the amorphous network of $SiO_2$ molecules [26, 27].

It is therefore of considerable interest to study the temperature dependence of the dissipation in ion-beam-sputtered $Ta_2O_5$ doped with $TiO_2$, here also an amorphous solid. We have chosen to study a single layer of doped $Ta_2O_5$ to allow investigation of its behaviour independent of the $SiO_2$ layers with which it would be used in a multi-layer coating. Previous measurements of dissipation in an $SiO_2$-$Ta_2O_5$ multilayer coating by Yamamoto *et al.* did not show strong temperature dependent effects [28]. The tantala component of these coatings was not doped, and measurements were taken over a limited temperature range.

## 2. Experimental procedure

The ion-beam-sputtered film under study (85.5 % Ta and 14.5 % Ti cation concentration) was applied to a thin silicon cantilever substrate of low mechanical loss $(47.7 \pm 0.5)\,\mu m$ thick and 34 mm long. A 30 nm thick thermal oxide layer was previously grown on the cantilever to allow the coating to adhere. The cantilever was fabricated from an n-type, antimony-doped single-crystal Si wafer (resistivity 0.005-0.25 $\Omega$ cm) by a hydroxide etch, with a thicker clamping block left at one end of the cantilever (see Figure 1) to reduce frictional energy loss into the clamp used to support the cantilever [29, 30].

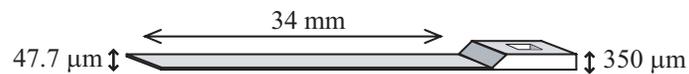

Figure 1. Schematic diagram of one of the silicon cantilevers used.

A second cantilever, nominally identical to the coated sample, underwent the same oxidisation so that the only difference between the two cantilevers was the presence of the doped-tantala layer. Assuming that all other sources of loss [31] are the same for the two samples, the dissipation in the coating layer can be calculated from the difference in the mechanical loss of the coated and uncoated cantilevers.

## 2.1 Method

Each cantilever was held horizontally by a stainless steel clamp within a cryostat, as shown in Figure 2. To ensure that gas damping effects were negligible the measurements were taken at a pressure of below $2 \times 10^{-6}$ mbar [31]. A resistive heater mounted on the clamp was used to control the temperature of the clamp and cantilever. The temperature was measured using a silicon diode sensor mounted within the clamp directly below the cantilever.

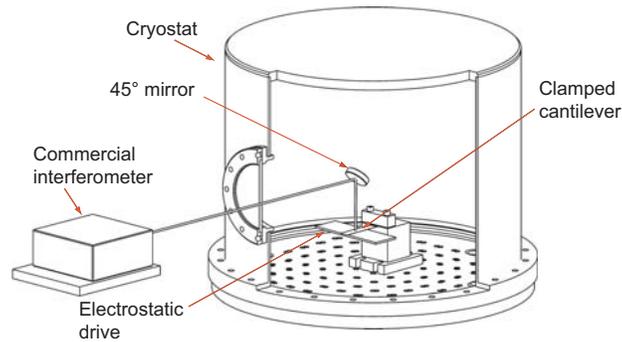

Figure 2. A schematic diagram of the experimental setup used, showing a sample clamped cantilever within the cryostat and the laser interferometer readout system.

The bending modes of each cantilever were excited in turn using an electrostatic drive. Once a mode had been suitably excited the drive plate was grounded and the free amplitude decay of the resonant motion monitored using a SIOS SPS-120/500 laser interferometer. The mechanical loss was calculated by fitting an exponential curve to the measured amplitude ringdown. The experimental design and technique used are described in detail in [31] and details of the cryostat can be found in [32].

## 3. Experimental Results

Figure 3 shows the measured mechanical loss of the first three bending modes of both the coated and control samples between 10 K and 292 K.  Each point is the average loss calculated from at least three ringdown measurements and has a standard error of typically less than 3%. Error bars have been omitted for clarity.

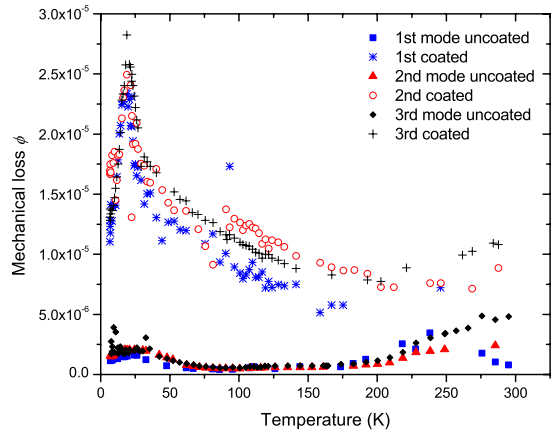

Figure 3. Temperature dependence of the measured mechanical loss of the coated and uncoated cantilevers for the 1$^{st}$ , 2$^{nd}$ and 3$^{rd}$ resonant modes at 55 Hz, 350 Hz and 989 Hz.

Each of the modes of the coated cantilever has a dissipation peak at approximately 20 K. The first two modes show some evidence of a second, smaller peak in the dissipation at approximately 90 K (see Figure 3). The loss of the third mode has a very similar trend, increasing steadily with decreasing temperature to a well defined peak at ~20 K, with no evidence of a peak at 90 K. It seems likely that this 90 K 'peak' observed in the first two modes was not intrinsic to the sample and instead was due to a temperature dependent coupling to the clamping structure [31], as discussed below for measurements of the uncoated cantilever around 200 K.

For each mode, the loss of the uncoated sample decreases to a plateau between approximately 175 K and 110 K, below which the loss rises to a relatively small broad peak at ~ 25 K. Above 200 K, the loss of the second and third modes of the uncoated sample follow the trend of the expected thermo-elastic loss of the cantilever [31]. The loss of the first mode, however, is significantly higher than the calculated thermo-elastic loss, and begins to decrease at temperatures above 230 K. A piezo-electric transducer attached to the clamp showed some evidence of low frequency motion in the clamp during ring-down measurements around 200 K, suggesting that as has been observed previously [31], energy loss into the clamping structure is responsible for the high losses measured for this mode at these temperatures.

It should be noted that the peak in the dissipation of the coated sample at 20 K is well defined for all of the modes measured, and that no evidence of energy coupling to the clamping structure was observed around this peak.

## 4. Coating loss analysis

The total measured loss in the coated cantilever can be described by the following equation [8]:

$$\phi(\omega_0)_{coated-sample} = \phi(\omega_0)_{substrate} + \frac{E_c}{E_s}\phi(\omega_0)_{coating} \tag{1}$$

where $\phi(\omega_0)_{coated-sample}$ is the mechanical loss of the coated cantilever, $\phi(\omega_0)_{substrate}$ the loss of the uncoated cantilever, $\phi(\omega_0)_{coating}$ the loss of the coating layer and $E_c/E_s$ is the ratio of the energy stored in the coating layer to that stored in the cantilever. Where the coating is thin in comparison to the substrate, this energy ratio is given by $E_c/E_s = 3Y_c t / Y_s a$ [33], where $Y_c$ and $Y_s$ are the Young's modulus of the coating and the substrate respectively, $t$ is the thickness of the coating and $a$ is the thickness of the substrate. Thus the loss of the coating is given by:

$$\phi(\omega_0)_{coating} = \frac{Y_s a}{3Y_c t}\left(\phi(\omega_0)_{coated-sample} - \phi(\omega_0)_{substrate}\right) \tag{2}$$

The Young's modulus of Ta$_2$O$_5$ was taken to be (140 ± 15) GPa [34] at room temperature. Assuming that the temperature dependence of Young's modulus is typical of other amorphous oxides [36], then its effect on the calculation over the temperature range studied here is negligible. Previous measurements have shown [15] that surface and interface effects have no significant contribution to the total coating loss: thus $\phi(\omega_0)_{coating}$ is treated as a uniform loss throughout the coating layer.

Equation 2 was used to calculate the mechanical loss of the doped tantala layer for each of the modes of the cantilever. As shown in Figure 4, the coating loss was generally found to steadily increase with decreasing temperature, from approximately $2\times10^{-4}$ at 292 K to a peak of approximately $1\times10^{-3}$ at ~20 K. This peak in the doped tantala loss is present for each of the modes measured. The temperature at the point of maximum dissipation increases with the mode frequency. For clarity, the loss of each of the modes at temperatures around the low temperature peak is shown in more detail in Figure 5, which also shows data measured for the next two bending modes of the sample.

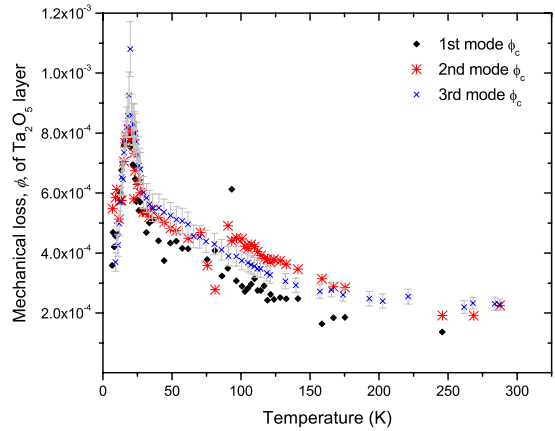

Figure 4. Temperature dependence of the loss of the doped $Ta_2O_5$ coating.

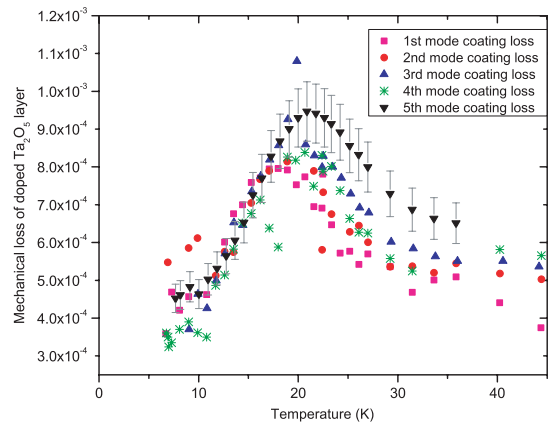

Figure 5. The mechanical loss peak measured in the doped $Ta_2O_5$ coating. Note the temperature at which the peak occurs increases with increasing mode frequency. For clarity, the error bars are only shown for the 5$^{th}$ mode: these are typical of the errors in all of the points.

If it is assumed that the dissipation peak has the form of a Debye peak, then the loss $\phi(\omega)$ can be expressed as:

$$\phi(\omega) = \Delta \frac{\omega\tau}{1+(\omega\tau)^2} \, , \tag{3}$$

where $\Delta$ is a constant related to the magnitude of the dissipation [36]. For a thermally activated dissipation process the characteristic time $\tau$ is given by the Arrhenius equation:

$$\tau^{-1} = \tau_0^{-1} e^{-\frac{E_a}{k_B T}} \, , \tag{4}$$

where $\tau_0^{-1}$ is the rate constant of the dissipation mechanism, $E_a$ is the activation energy and $k_B$ is Boltzmann's constant [23]. At the dissipation peak, $\omega\tau = 1$ [36] and thus we can write $\omega = \tau_0^{-1} \exp(-E_a/k_B T_{peak})$ and so

$$\ln(\omega) = \ln(\tau_0^{-1}) - \frac{E_a}{k_B T_{peak}} \, . \tag{5}$$

Thus a graph of the natural logarithm of the angular mode frequency against $1/T_{peak}$ yields a straight line of slope $E_a/k_B$, as shown in Figure 6. The activation energy is $(42 \pm 2)$ meV and the rate constant is $3.3 \times 10^{14}$ Hz. The activation energy of the well-known low temperature dissipation peak in fused silica is approximately 44 meV [27]. We deduce that the mechanism responsible for the peak observed here may be thermally activated transitions of oxygen atoms between two states in a double well potential, analogous to the mechanism in fused silica.

Following the analysis of [37] the experimental values of $E_a$ can be used to predict the shape of the loss peak. This theoretical peak is substantially narrower than the observed experimental peak which is characteristic of strongly interacting or disordered systems in which there is a range of values $E_a$ [37]. Since the $Ta_2O_5$ layer has an amorphous structure (see Figure 7), we postulate that this range of values of $E_a$ may be related to a distribution of the $Ta_2O_5$ bond angles.

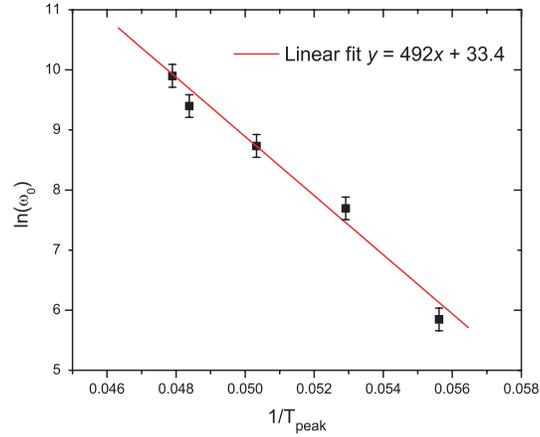

Figure 6. Plot of $\ln(\omega_0)$ against $1/T_{peak}$.

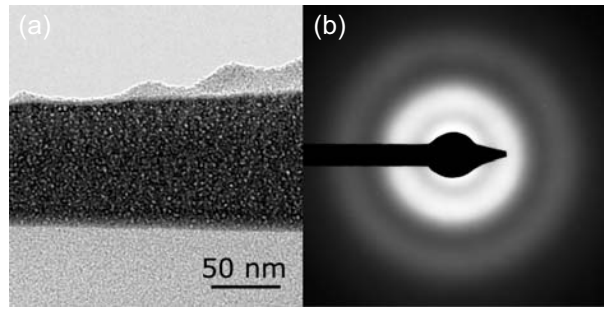

Figure 7. (a) Bright field TEM image of a tantala layer in a $SiO_2$ / $Ta_2O_5$ multilayer coating, where the $Ta_2O_5$ layers are doped with 8 % $TiO_2$ (b) convergent beam electron diffraction from a pure $Ta_2O_5$ layer showing just diffuse rings of intensity, confirming that this layer is amorphous.

It has been shown that heat treatment of fused silica can significantly reduce the observed mechanical loss [38]. Heating is believed to alter the bonding structure in the material, thus narrowing the distribution of potential barrier heights in the double well systems. We therefore believe that extended heat treatment at temperatures below 650°C, at which point $Ta_2O_5$ films are known to crystallise, should reduce the width of the dissipation peak, resulting in lower dissipation, and thus thermal noise, at temperatures above and below the peak.

## 5. Conclusions

We have identified an important dissipation mechanism, analogous to the double well potential loss mechanism observed in fused silica, in the titania-doped tantala coating studied. Following experience with silica it should be possible to reduce the width of the dissipation peak and hence the associated thermal noise at temperatures above and below that of the peak by suitable thermal annealing techniques. This will be of high significance for a range of precision measurements in fundamental physics.

**Acknowledgements**

This work was supported by STFC and the University of Glasgow, the NSF through grants NSF PHY-0140297 (Stanford), NSF 0601135 (Embry-Riddle) and NSF PHY-0355118 (Hobart and William Smith Colleges) and the German science foundation (DFG) under SFB Transregio 7. We also wish to thank the ILIAS Strega project and Leverhulme Trust for support. The LIGO Observatories were constructed by Caltech and MIT with funding from the NSF under cooperative agreement PHY-9210038. The LIGO Laboratory operates under PHY-0107417.